\documentclass[usegraphicx,usenatbib]{mn2e}

\usepackage{mathptmx}
\usepackage{amsmath}
\usepackage{times}

\newcommand{\Mpc}{\rm\thinspace Mpc}
\newcommand{\kpc}{\rm\thinspace kpc}

\newcommand{\km}{\rm\thinspace km}

\newcommand{\cm}{\rm\thinspace cm}

\newcommand{\pcmcu}{\hbox{$\cm^{-3}\,$}}

\newcommand{\Gyr}{\rm\thinspace Gyr}
\newcommand{\s}{\rm\thinspace s}
\newcommand{\ks}{\rm\thinspace ks}

\newcommand{\K}{\rm\thinspace K}

\newcommand{\keV}{\rm\thinspace keV}

\newcommand{\kmps}{\hbox{$\km\s^{-1}\,$}}

\newcommand{\kmpspMpc}{\hbox{$\kmps\Mpc^{-1}$}}

\newcommand{\Zsun}{\hbox{$\thinspace \mathrm{Z}_{\odot}$}}

\newcommand{\psqcm}{\hbox{$\cm^{-2}\,$}}
\newcommand{\pcmsq}{\hbox{$\cm^{-2}\,$}}

\newcommand{\amin}{\rm\thinspace arcmin}

\newcommand{\asec}{\rm\thinspace arcsec}

\newcommand{\cts}{\rm\thinspace counts}

\title[Direct Spectral Deprojection] {Direct X-ray Spectral
  Deprojection of Galaxy Clusters} \author[H.R. Russell et
al.]{H.~R. Russell\thanks{E-mail: hrr27@ast.cam.ac.uk},
  J.~S. Sanders and A.~C. Fabian\\
\footnotesize
Institute of Astronomy, Madingley Road, Cambridge CB3 0HA\\
}

\begin{document}

\maketitle

\begin{abstract}
  Temperature, density and abundance profiles of the hot intracluster
  medium (ICM) are important diagnostics of the complex interactions
  of gravitational and feedback processes in the cores of galaxy
  clusters.  Deprojection of X-ray data by methods such as
  \textsc{projct}, which are model dependent, can produce large and
  unphysical oscillating temperature profiles.  Here we validate a
  deprojection routine, Direct Spectral Deprojection
  (\textsc{dsdeproj}; \citealt{SandersFabian07}), showing that it
  solves some of the issues inherent to model-dependent deprojection
  routines.  \textsc{dsdeproj} is a model-independent approach,
  assuming only spherical symmetry, which subtracts projected spectra
  from each successive annulus to produce a set of deprojected
  spectra.
\end{abstract}


\section{Introduction}

X-ray observations of many low-redshift galaxy clusters find strongly
peaked emission in the centre from relatively high density gas with a
radiative cooling time below $\sim1\Gyr$.  Without a compensating
source of heat, this gas should cool rapidly causing a drop in the
central pressure and a subsequent slow inward flow of the overlying
gas known as a cooling flow (\citealt{Fabian94}).  This cooling flow
gas should continue to cool to low temperatures.  However, recent high
resolution X-ray spectroscopy was unable to find the emission
signatures of gas cooling below about $\sim1\keV$
(\citealt{Peterson03}; \citealt{Kaastra04};
\citealt{PetersonFabian06}).  In addition, the cool gas masses implied
by a cooling flow exceeded the sum of the observed gas masses
(eg. \citealt{EdgeFrayer03}) and inferred star formation rates by an
order of magnitude (\citealt{Johnstone87}; \citealt{HicksMushotzky05};
\citealt{Rafferty06}, \citealt{ODea08}).  A heating mechanism is
therefore required to prevent significant amounts of cooling from
occurring (see \citet{McNamaraNulsen07} for a review).

High spatial resolution images of cluster cores from the \emph{Chandra
  X-ray Observatory} have revealed huge cavities, shock fronts and
cold fronts in the ICM.  Cold fronts are sharp boundaries between
regions of different temperature and density likely caused by gas
sloshing in cluster cores.  The X-ray cavities, or bubbles, in
the hot gas have been produced by the interaction of jets from
the central AGN with the surrounding ICM (eg. \citealt{FabianPer03};
\citealt{FormanM8705}; \citealt{FabianPer06};
\citealt{McNamaraNulsen07}).  These cavities provide a direct and
relatively reliable means of measuring the energy injected by the
central radio source.  However, whilst AGN heating has been shown to
be capable of energetically balancing cooling losses
(\citealt{Birzan04}; \citealt{Rafferty06}; \citealt{DunnFabian06};
\citealt{McNamara06}), the exact mechanisms by which the energy is
transported and dissipated and how the balance is achieved are still
unclear.

Structures such as shock fronts and cold fronts provide tools to study
the microphysics and transport processes at work in the ICM.  These
discontinuities allow the measurement of the bulk velocities of the
gas in the plane of the sky and the stability of these features provide
constraints on thermal conduction, the formation of magnetic draping layers and
the strength of hydrodynamic instabilities.  Detailed studies of shocks and
cold fronts in clusters have shown that transport processes in the ICM can
be easily suppressed (\citealt{Ettori00}; \citealt{Vikhlinin01}). 

In order to study shocks, cavities and cold fronts in the cluster core,
and to relate the properties of cool core gas to the strength of the
AGN feedback, we require a reliable method for extracting the core gas
properties.  A spectrum extracted from the cluster centre on the plane
of the sky corresponds to a summed cross-section with a range of
spectral components from the core to the cluster outskirts.  The
spectral properties at any point on the plane of the sky are the
emission-weighted superposition of radiation from all points along the
line of sight through the cluster.  These superimposed contributions
from the outer cluster layers can be subtracted off by making an
assumption about the line of sight extent, usually assuming the
cluster is spherical, and deprojecting the emission to produce
deprojected radial profiles in temperature, density and metallicity.

The most commonly used deprojection routine is \textsc{projct} in the
X-ray spectral fitting package \textsc{xspec} (\citealt{Arnaud96}).
\textsc{projct} takes spectra from a series of concentric annuli and
fits each one with a set of models to account for all the layers of
projected emission plus a model for the innermost emitting region.
However, \textsc{projct} has been found to be unstable to changes in
the radial binning and produce oscillating temperature profiles that
do not relate to physical changes in the gas temperature
(eg. \citealt{FabianPer06}).  In section \ref{sec:projct}, we use
simulated clusters to show that multiphase gas can be responsible for
the unstable \textsc{projct} deprojections.

In section \ref{sec:DSDeproj}, we describe a spectral deprojection
routine, \textsc{dsdeproj} and compare the results with
\textsc{projct} for simulated clusters.  In section \ref{sec:sample},
we apply both \textsc{dsdeproj} and \textsc{projct} to a set of three
nearby galaxy clusters which contain shocks, cavities and knots of
cool gas in the core.  Finally, in section \ref{validation}, we show
that \textsc{dsdeproj} produces stable radial profiles for simulated
clusters that have multiple temperature components, sharp breaks in
temperature and density or that are extended along the line of sight.

We used \textsc{xspec} version 11.3.2 (\citealt{Arnaud96}) for all
spectral fitting.  The \textsc{mekal} spectral model
(\citealt{Mewe85}, \citeyear{Mewe86}; \citealt{Kaastra92};
\citealt{Liedahl95}) and \textsc{projct} deprojection model used here
are the defaults available in that version of \textsc{xspec}.
Photoelectric absorption was modelled with the \textsc{phabs} model
(\citealt{Balucinska92}).  Abundances were measured assuming the
abundance ratios of \citet{AndersGrevesse89}.

We assume $H_0 = 70\kmpspMpc$, $\Omega_m=0.3$ and
$\Omega_\Lambda=0.7$.  All errors in fit parameters are $1\sigma$ unless
otherwise noted.

\section{Projct}
\label{sec:projct}
\textsc{projct} fits spectra extracted from a series of concentric annuli
simultaneously, to account for each projected component.  Each
projected spectrum is fitted by one or more components described by a set
of parameters, such as temperature and density.  Assuming spherical
symmetry to calculate suitable geometric factors (\citealt{Kriss83}),
the projected sum of the components along line of sights are fitted to
each spectrum simultaneously.  \textsc{projct} reads the inner and
outer radii of each annulus from FITS header items to calculate the
volume of each corresponding shell.  

\citet{Johnstone05} tested the \textsc{projct} model on synthetic,
single-temperature cluster spectra and found that the fitted
temperature and density profiles agreed with the true profiles.
However, the deprojected temperature profiles generated by
\textsc{projct} have been found to oscillate between values separated
by several times the uncertainties on the values (eg. for the Perseus
cluster, \citealt{FabianPer06}).  This oscillation can disappear if
different sized annuli are used, indicating that this is not related
to physical changes in the cluster properties.  \citet{FabianPer06}
suggest that this instability may be caused by multiphase gas or
deviation from spherical geometry.

\subsection{Simulated Clusters}
In order to thoroughly test our deprojection routines, we produced a
set of simulated clusters with a range of properties including
two-temperature components, sharp breaks in density or temperature and
an elongated geometry.  The generic simulated cluster was based on a
$500\ks$ observation of the Perseus cluster, with redshift $z=0.0183$ and
Galactic absorption $n_\mathrm{H}=0.1 \times 10^{22} \pcmsq$.  The
radial dependence of the deprojected parameters were given by the
analytical approximations:

\begin{equation}
n_e=\frac{3.9\times10^{-2}}{\left(1+\left(r/80\kpc\right)^2\right)^{1.8}}+\frac{4.05\times10^{-3}}{\left(1+\left(r/280\kpc\right)^2\right)^{0.87}} \;\;\;\pcmcu
\label{n_e}
\end{equation}
\begin{equation}
T = \frac{7\times\left(1+\left(r/100\kpc\right)^3\right)}{\left(2.3+\left(r/100\kpc\right)^3\right)} \;\;\;\keV
\label{T}
\end{equation}
\begin{align}
\nonumber Z &= 0.35 + 0.0139\left(\frac{r}{\kpc}\right) -
0.000243\left(\frac{r}{\kpc}\right)^2 \\ 
&\;\;\;\;\;\;\;\;\; +1.031\times10^{-6}\left(\frac{r}{\kpc}\right)^3
\Zsun \;\;\; \mathrm{for} \: r < 121\kpc \\
& = 0.3 \Zsun \;\;\; \mathrm{for} \: r > 121\kpc
\label{Z}
\end{align}

The radial dependence of the density was determined by \citet{Churazov03}, where
the second term, which describes the distribution on larger scales, was taken
from \citet{JonesForman99}.  The expression for the gas temperature
distribution was also taken from \citet{Churazov03}.  An approximate
fit to the observed abundance profile of the Perseus cluster was used
for the abundance distribution. 

The \textsc{fakeit} command and a
\textsc{phabs}(\textsc{mekal}$+$\textsc{mekal}) model in
\textsc{xspec} were used to create artificial spectra for the
two-temperature galaxy cluster with the parameters set according to the
supplied equations and then folded through suitable response files.
The input parameters for the second \textsc{mekal} component were also
determined by equations \ref{n_e}--\ref{Z}, but the temperature and
density at each radius were halved.  The projected spectra were
generated by stepping through the cluster radius on the sky and, for
each radius, summing all the artificial spectra along the line of
sight.  The final simulated projected spectra were grouped into bins
with a minimum of 50 counts, allowing the use of $\chi^2$ statistics.

This same simulated dataset was used throughout sections
\ref{sec:projct} and \ref{sec:DSDeproj} to allow a consistent
interpretation of the deprojection methods.  The enhanced metallicity
in the second annulus from the centre, which for these noisy core
spectra results in lower temperature and density best-fitting values, and
the fluctuations in the profiles from $50-100\kpc$ are particular to
this simulated cluster (eg. Figure \ref{fig:Fakecluster_2compprojct}).
Deprojections of other simulated clusters do not also display these
specific features.

\subsection{Single-temperature Model}
\begin{figure}
  \includegraphics[width=\columnwidth]{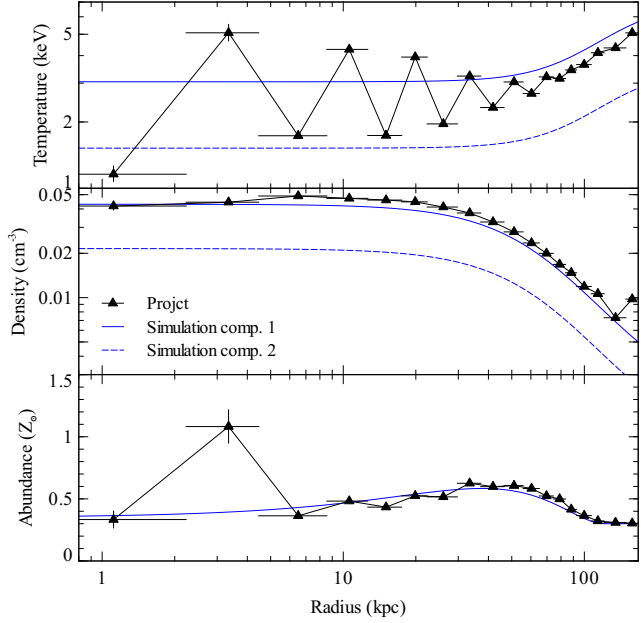}
  \caption{Deprojected temperature (top), electron density (centre)
    and metallicity (bottom) profiles for a two-temperature simulated
    cluster.  The single-temperature \textsc{projct} fit to the data
    is overlaid on the true profiles.}
  \label{fig:Fakecluster_projct}
\end{figure}

We fitted these two-temperature synthetic spectra with an absorbed
single-temperature \textsc{mekal} thermal model and \textsc{projct} in
\textsc{xspec 11.3.2}.  The redshift and Galactic hydrogen column
density were fixed to the simulated cluster values $z=0.0183$ and
$n_\mathrm{H}=0.1 \times 10^{22}\pcmsq$.  The temperature, metallicity and model
normalization (relating to density) parameters were allowed to vary
and the resulting profiles are shown overlaid on the true profiles in
Figure \ref{fig:Fakecluster_projct}.  The single-temperature
\textsc{projct} profile appears to bounce between the two temperature
components suggesting that \textsc{projct} tends to account for one
temperature component in one shell and the other component in a
neighbouring shell.  If we assume that the single-temperature spectral
model is a good fit to the data, a very misleading result is produced.

As an aside, we note that the bounce in the outermost annulus of the density
profile (Figure \ref{fig:Fakecluster_projct}) is caused by a different
effect.  The emission extends beyond the outermost annulus of each
cluster we analysed so that the emitting volume associated with this
annulus is too small.  This causes an overestimate of the projection
on to the next annulus in, so that the fit to that annulus
underestimates its brightness.  However, the effect is minimal for the
annuli beyond the outer two because the steep surface brightness
profile limits the amount of projection.

\subsection{Fixing Parameters in \textsc{projct}}
\begin{figure}
  \includegraphics[width=\columnwidth]{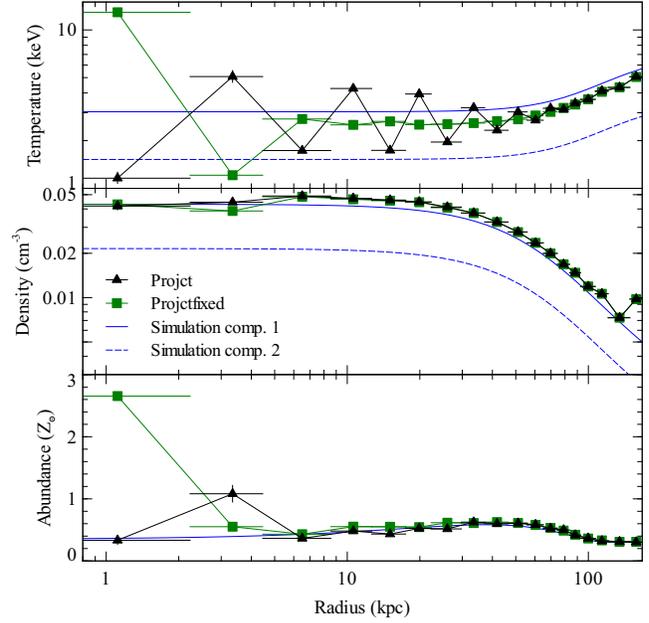}
  \caption{Deprojected temperature (top), electron density (centre)
    and metallicity (bottom) profiles for a two-temperature simulated
    cluster.  The standard single-temperature \textsc{projct} fit and
    the single-temperature \textsc{projctfixed} fit are overlaid on
    the true cluster profiles.  There are no vertical error bars for
    the \textsc{projctfixed} model (green line)}
  \label{fig:Fakecluster_fixedpars}
\end{figure}

The oscillation in the temperature profile can sometimes be alleviated
by fitting the annuli sequentially from the outside and freezing the
parameters of components in the outer shells before fitting spectra
from shells inside them (\textsc{projctfixed}; see
eg. \citealt{Sanders04}).  This prevents the poorly modelled spectra
near the centre from affecting the results in outer annuli.  However,
this method underestimates the uncertainties on the model parameters
because the outer shell uncertainties are not included.  Figure
\ref{fig:Fakecluster_fixedpars} shows that freezing the outer shell
parameters does remove some of the oscillation in the outer shells but
that this approach produces particularly poor fits for the innermost
annuli.

\subsection{Two-temperature Model}
A two-temperature \textsc{mekal} model can be fitted to the data if
there are sufficient counts to produce well-constrained parameters.
We reduced the number of free model parameters in each annulus
by tying the abundance parameters together, leaving the two
temperature parameters and the normalizations free.  The redshift and Galactic
hydrogen column density were fixed to the simulated cluster values
$z=0.0183$ and $n_\mathrm{H}=0.1 \times 10^{22}\pcmsq$.   

\begin{figure}
  \includegraphics[width=\columnwidth]{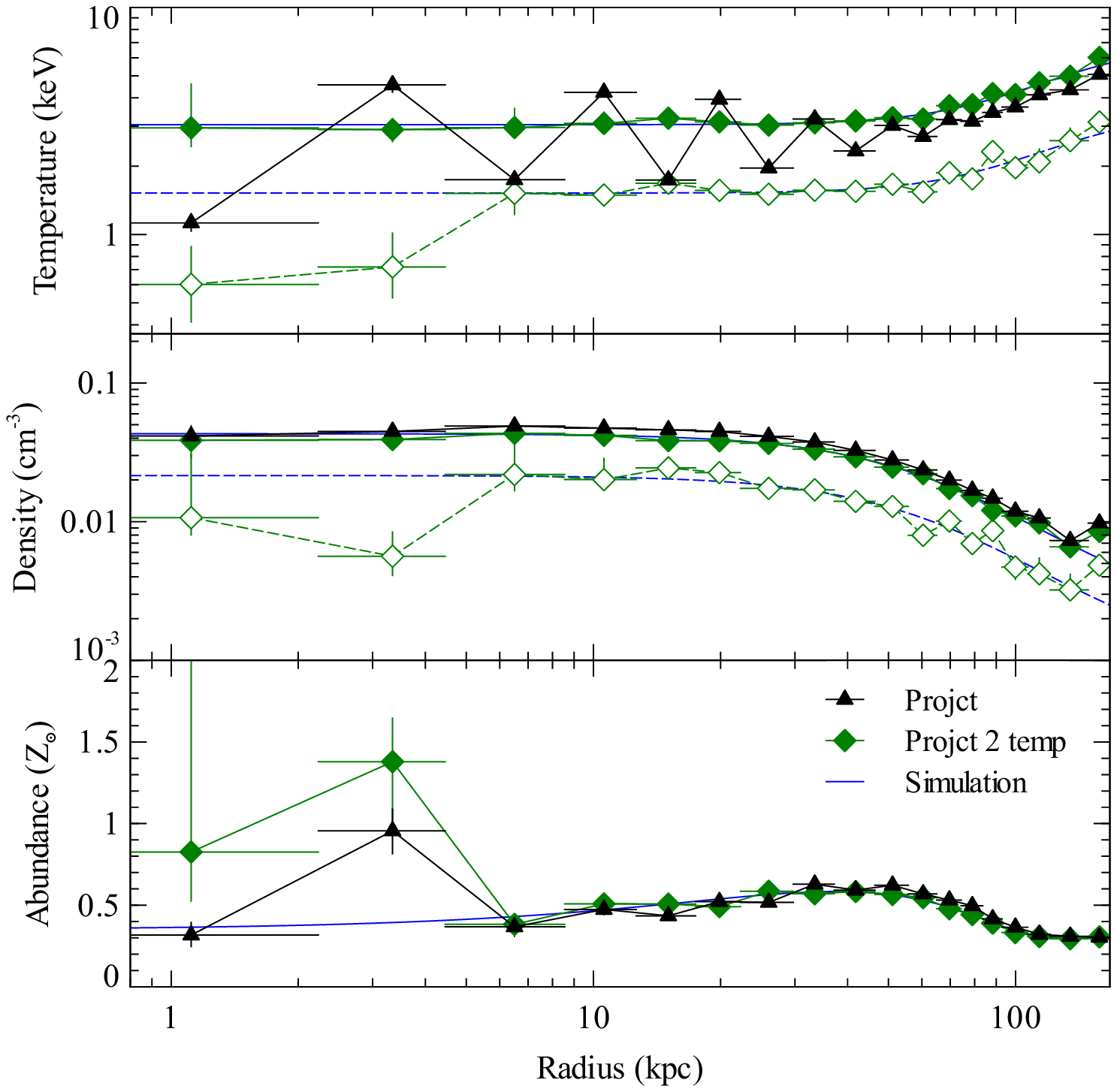}
  \caption{Deprojected temperature (top), electron density (centre)
    and metallicity (bottom) profiles for a two-temperature simulated
    cluster with lines indicating the two spectral components of the
    true cluster (solid and dashed blue), single-temperature
    \textsc{projct} fit (black triangles), two-temperature \textsc{projct} fit
    (solid and open green diamonds).}
  \label{fig:Fakecluster_2compprojct}
\end{figure}

Figure \ref{fig:Fakecluster_2compprojct} shows that a two-temperature
model provides a much better fit to the simulated profiles with
$\chi^2_\nu=1.00$ compared to $\chi^2_\nu=4.56$ for the single-component
model.  The two-temperature model closely follows the true cluster
profiles except for the two innermost annuli.  The spectra from the
inner annuli contain a large quantity of projected emission which,
when subtracted off, leaves particularly noisy deprojected spectra for
those shells. 

In summary, we have shown that assuming a single-component
\textsc{projct} model for a cluster with two spectral components,
produces an oscillating temperature profile with values that are unrelated to
physical changes in the cluster gas.  Similar deprojection methods
which assume a spectral model for the deprojection will suffer from
the same problem.  To provide a solution to this issue, we present and
validate a new deprojection routine, \textsc{dsdeproj}, which assumes only
spherical geometry.

\section{\textsc{dsdeproj}}
\label{sec:DSDeproj}
\textsc{dsdeproj} (\citealt{SandersFabian07}) is a model-independent
deprojection method which, assuming only spherical symmetry, uses a
purely geometrical procedure to subtract off the projected emission in
a series of shells (similar to \citealt{Nulsen95}).  The resulting deprojected
spectra can then be fitted with single or multiple temperature models
in \textsc{xspec} to produce deprojected profiles in temperature,
density and metallicity.  

\textsc{dsdeproj} takes as inputs: spectra extracted from a series of
concentric annuli in a sector of the cluster and suitable blank-sky
backgrounds.  \textsc{dsdeproj} then performs the following steps:

\begin{enumerate}
\item Subtract the equivalent background spectrum from each of the foreground
  cluster spectra
\item Start at the spectrum from the outermost annulus: divide the
  count rate in each spectral energy bin by the emitting volume,
  assuming it was emitted from a section of a spherical shell,
  (geometric factors from \citealt{Kriss83}) to give a spectrum per
  unit volume
\item Scale the spectrum per unit volume from the outermost annulus by
  the volume projected onto the neighbouring inner annulus, and
  subtract this from the count rate in each spectral bin of that
  annulus
\item Calculate a new count rate per unit volume in each spectral bin
  for this inner shell
\item By moving inwards to each successive annulus, we can subtract
  off the projected emission of each outer shell from the inner shells
  and produce a set of deprojected spectra
\end{enumerate}

We used a Monte Carlo technique to calculate the uncertainties in the
count rate of each spectral channel in each spectrum.  Each of the
input foreground cluster and background spectra are binned to have 200
counts per spectral channel so that Gaussian errors can be assumed.
The deprojection process is repeated 6000 times, each time creating
new input foreground cluster and background spectra by simulating
spectra drawn from Gaussian distributions based on the initial spectra
and their uncertainties.  The output spectra are the median output
spectra calculated in this process.  The 1$\sigma$ errors on the count
rates in each spectral channel are calculated from the 15.85 and 84.15
percentile spectra.

\subsection{Single-temperature Model}
\begin{figure}
  \includegraphics[width=\columnwidth]{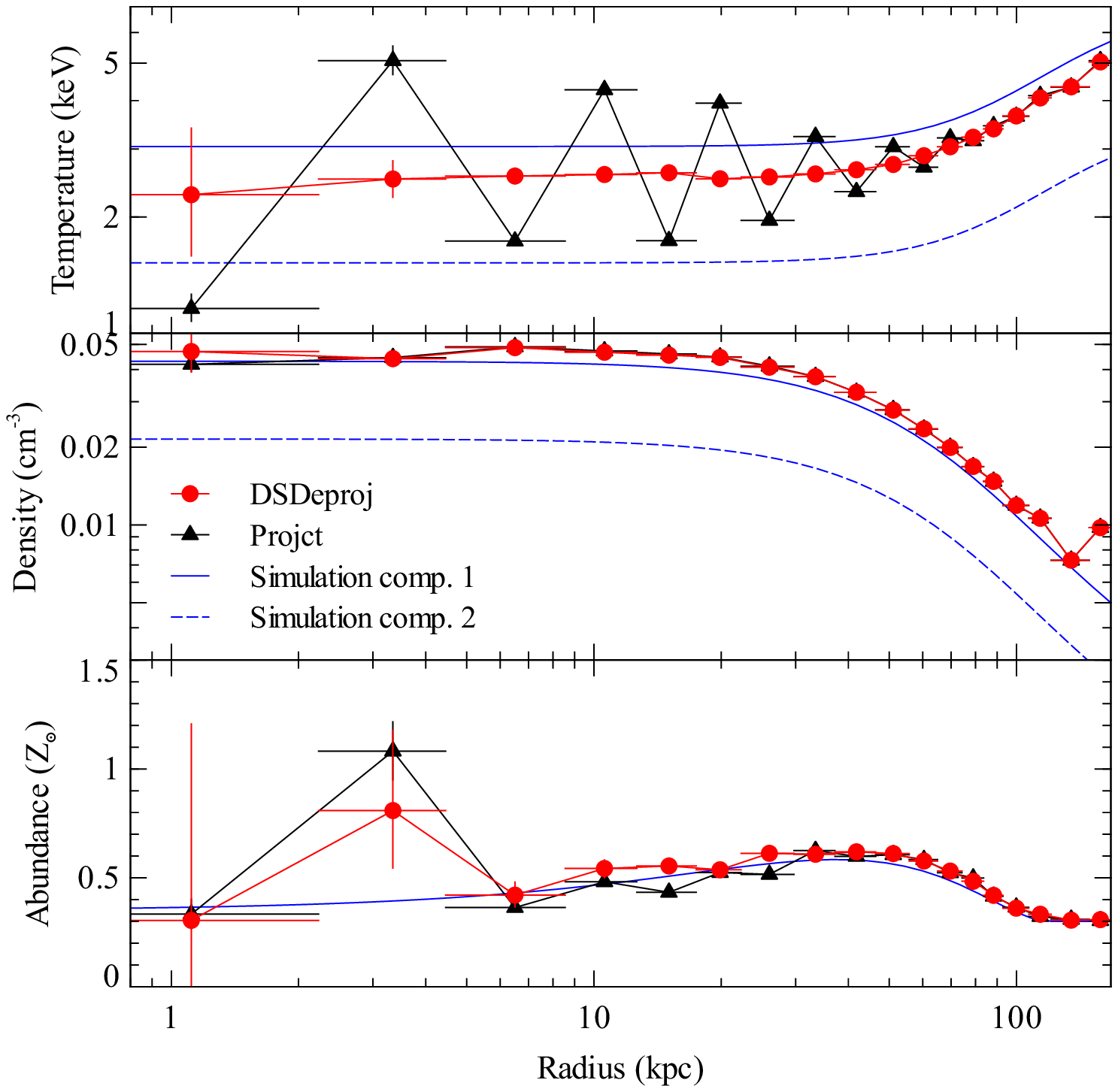}
  \caption{Deprojected temperature (top), electron density (centre)
    and metallicity (bottom) profiles for a two-temperature simulated
    cluster.  The single-temperature \textsc{projct} and
    \textsc{dsdeproj} results are overlaid on the true cluster
    profiles.}
  \label{fig:Fakecluster}
\end{figure}

Figure \ref{fig:Fakecluster} shows the deprojection of the
two-temperature simulated cluster spectra with \textsc{dsdeproj} and
the previous result from \textsc{projct}.  \textsc{dsdeproj} produces
a smooth temperature profile that is the average of the two separate
components (weighted by emission).  The errors in the parameters
increase towards the cluster centre as each subsequent annulus has an
increasing amount of projected emission which must be subtracted off.
This is particularly apparent for the central two radial bins.

\subsection{Two-temperature Model}
\begin{figure}
  \includegraphics[width=\columnwidth]{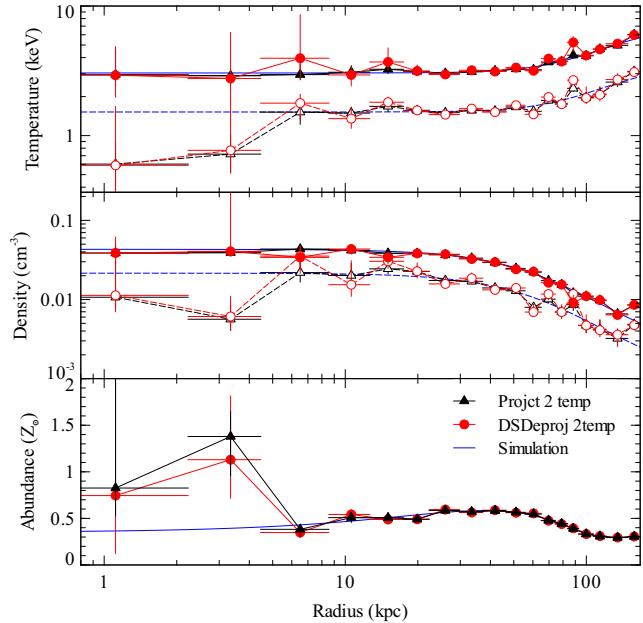}
  \caption{Deprojected temperature (top), electron density (centre)
    and metallicity (bottom) profiles for a two-temperature simulated
    cluster.  The two-temperature \textsc{projct} (black solid and
    open triangles) and \textsc{dsdeproj} results (red solid and open circles)
    are overlaid on the true cluster profiles (blue solid and dashed
    lines).}
  \label{fig:Fakecluster_2compallfree}
\end{figure}

The deprojected spectra produced by \textsc{dsdeproj} were also fitted
with an absorbed two-temperature \textsc{mekal} model to check that
the two components of the simulated cluster are correctly reproduced
(Figure \ref{fig:Fakecluster_2compallfree}).  We compared the result
with that of \textsc{projct} by leaving the two temperature parameters
and the normalization free and tying the abundance parameters for the
two components together.  The redshift and Galactic hydrogen column
density were fixed to the simulated cluster values $z=0.0183$ and
$n_\mathrm{H}=0.1 \times 10^{22}\pcmsq$.  We found that both \textsc{dsdeproj} and
\textsc{projct} recovered the simulated cluster profiles with a two
temperature spectral model.

A detailed validation of the \textsc{dsdeproj} routine for a series
of simulated clusters is given in section \ref{validation}.   

\section{Deprojection of a Sample of Galaxy Clusters}
\label{sec:sample}
We applied \textsc{dsdeproj} and \textsc{projct} to \emph{Chandra}
archive observations of three nearby ($z<0.1$) galaxy clusters:
Perseus, Hydra A and Abell 262.  These clusters each have deep archive
observations with total exposure times of $890\ks$, $200\ks$ and
$110\ks$, respectively.  Deep observations of nearby bright clusters
yield a large number of counts for each cluster which ensures a
detailed deprojection with a large number of fine bins and a
significant detection of multiple temperature components.

\subsection{Data Reduction}
The data were analysed using CIAO version 4.0 beta 2 with CALDB
version 3.4.1 provided by the \emph{Chandra} X-ray Center (CXC).  The
level 1 event files were reprocessed to apply the latest gain and
charge transfer inefficiency correction and then filtered for bad
grades.  The improved background screening provided by VFAINT mode was
applied where available.  The background light curves of the resulting
level 2 events files were filtered to remove periods affected by
flares using the \textsc{CIAO} script \textsc{lc\_clean}.  Background
spectra were extracted from the blank-sky background data sets
available from the CXC and cleaned using the same method applied to
the cluster observations.  These background spectra were normalized so
that the count rate of the source and background observations matched
in the $9-12\keV$ band.

For each cluster, we identified a suitable sector for deprojection and
extracted a series of spectra in concentric annuli centred on the
surface brightness peak.  In the Perseus cluster we selected a sector
containing the weak shock analysed in \citet{Graham08}.  For Abell 262
we defined a sector which enclosed the half of the cluster where the
X-ray emission was most extended.  Spectra from Hydra A were extracted
from complete annuli.  Individual sources were identified using the
\textsc{wavdetect} algorithm in CIAO.  Sources were visually confirmed
on the X-ray image and excluded from the analysis.

Each cluster sector was divided up into a series of annuli to give a
minimum of 3000 counts in each deprojected spectrum.  This criterion
provides enough deprojected counts to allow a good spectral fit and
measurement of the temperature, density and abundance in each shell.
All spectra were analysed in the energy range $0.5-7\keV$ using
\textsc{xspec} version 11.3.2 and grouped with a minimum of 50 counts
per spectral bin.  Response and ancillary response files were
generated for each cluster spectrum, weighted according to the number
of counts between 0.5 and $7\keV$.  The cluster spectra were then
deprojected with \textsc{dsdeproj} and \textsc{projct} in
\textsc{xspec}.  In addition, we fitted the annuli in series using the
\textsc{projct} model with the parameters in the outer annuli fixed to
their best fit values (see Section \ref{sec:projct}).

For the single-temperature deprojections, an absorbed plasma \textsc{mekal}
model was fitted to each spectrum in \textsc{xspec}.  The redshift was
fixed to the values given in Table \ref{tab:obs} and the absorbing
column density was fixed to the Galactic values given by
\citet{Kalberla05}.  The temperature, metallicity and model
normalization were allowed to vary.

The \textsc{projct} two-component model fitted to these real clusters
was not readily able to find a minimum.  The large number of
parameters meant that the fit was poorly constrained and unable to
converge on a solution.  We exclude this model from our analysis of
the cluster sample.  The two-temperature \textsc{dsdeproj} model also produced
poorly constrained parameters when fitted to the cluster sample.  The
use of this model was restricted to the Perseus cluster, for which the
best quality data is available.  We used a set of 20 annuli with a
minimum of $190,000\cts$ per annulus to produce deprojected spectra
which could constrain a two-temperature fit in each shell.

\begin{table*}
\begin{minipage}[t]{\textwidth}
\centering
\caption{\emph{Chandra} observations included in this analysis.} 
\begin{tabular}{l c c c c c c}
\hline 
Cluster & Obs. ID & Aim Point & ACIS Mode & Total Exposure (ks) &
z & $n_\mathrm{H}$$^1$\\
\hline
Abell 262 & 7921 & ACIS-S & VFAINT & 110 & 0.0166 & 0.0567 \\
Hydra A & 4969, 4970 & ACIS-S & VFAINT & 200 & 0.0549 & 0.0468 \\
Perseus & 3209, 4289, 4946, 4947, 4948, & ACIS-S & FAINT & 890 &
0.0183 & 0.132 \\
 & 4949, 4950, 4951, 4952, & & & & & \\
 & 4953, 6139, 6145, 6146 & & & & & \\
\hline
\end{tabular}
\label{tab:obs}
\begin{tabular}{l}
$^1$Galactic absorption column density (in units of $10^{22}\psqcm$) adopted in this
paper (\citealt{Kalberla05}). \\
\end{tabular}
\end{minipage}
\end{table*}

\subsection{The Perseus cluster}

The data used for this analysis were first presented in the deep study
of the Perseus cluster (Abell 426) in \citet{FabianPer06}.  The total
good exposure time from the combined \emph{Chandra} observations is
$890\ks$.  This deep observation has revealed details of the complex
interaction between the central AGN and the surrounding ICM.
Depressions or cavities in the X-ray correspond to bubbles of
relativistic plasma that have been inflated by jets from the nucleus
and displaced the surrounding gas (\citealt{Boehringer93};
\citealt{FabianPer00}).  The outer bubbles are presumably from past
activity, having detached from the nucleus and risen buoyantly
outwards through the cluster core.  Shocks and cool gas around the
inner bubbles and ripples in the surrounding gas, interpreted as sound
waves, also provide a challenge for any deprojection routine.

\begin{figure}
  \centering
  \includegraphics[width=\columnwidth]{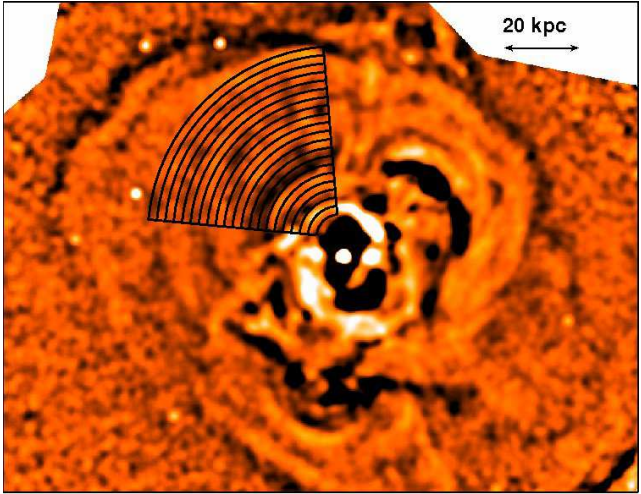}
  \caption{Unsharp mask image of the Perseus cluster made from the
    $0.3-7\keV$ band by subtracting an image smoothed with a Gaussian
    of width $10\asec$ from one smoothed by $2.5\asec$ and dividing by
    the sum of the two images.  The annuli used for deprojection have been overlaid in black.}
  \label{fig:Perseusimage}
\end{figure}

For our deprojection analysis we focused on a sector of the cluster core
containing a section of the weak shock that surrounds the inner bubbles
(Figure \ref{fig:Perseusimage}; \citealt{Graham08}).  Spectra were extracted from a series of 20
radial bins in this sector spaced $0.1\amin$ in radius.
In an analysis of a similar region, \citet{FabianPer06} found the
deprojected temperatures produced by \textsc{projct} to be unstable
depending on which radial bins were used.  They concluded that this may
be due to non-spherical geometry or a multiphase gas.

\citet{SandersFabian07} found significant variations in the model
absorption parameter, $n_\mathrm{H}$, across the image of the Perseus
cluster.  Most of this variation is likely due to the cluster's
proximity to the Galactic plane.  Leaving the $n_\mathrm{H}$ parameter
free for the spectral fits in this analysis did not produce any
significant differences in the radial profiles.

\begin{figure}
  \includegraphics[width=\columnwidth]{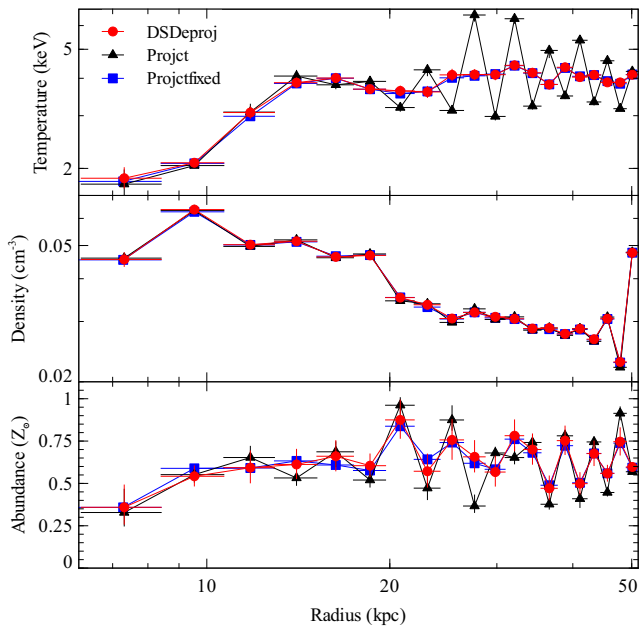}
  \caption{Deprojected temperature (top), electron density (centre)
    and metallicity (bottom) profiles for the Perseus cluster.  There
    are no vertical error bars for the \textsc{projctfixed} model (blue line).}
  \label{fig:Perseus_DSDeproj}
\end{figure}

Figure \ref{fig:Perseus_DSDeproj} shows the results of the
deprojection routines with single-temperature spectral models applied
to the sector of the Perseus cluster.  In agreement with
\citet{FabianPer06}, we found that \textsc{projct} produced an
unstable deprojected temperature profile with temperatures depending
on the positioning of the radial bins.  This oscillation, from
$\sim20-50\kpc$ in the top panel of Figure \ref{fig:Perseus_DSDeproj},
is caused by the multiphase gas in the Perseus cluster core.
\textsc{dsdeproj} produced a smooth temperature profile, revealing a
decline in temperature from $\sim4\keV$ to below $2\keV$.
\textsc{projctfixed} produced similar deprojected profiles although
this method does not correctly calculate errors on the parameters.


\textsc{projct} and \textsc{dsdeproj} produced similar stable
deprojected density profiles (middle panel of Figure
\ref{fig:Perseus_DSDeproj}).  The sharp increase in density in the
outermost radial bin for all three methods shows that there is still a
significant amount of projected emission in that bin.  The outer two
bins should therefore be excluded from any quantitative analysis.  The
two annuli around $11\kpc$ show a drop in both the temperature and
density, corresponding to a drop in the thermal pressure of the gas.
There is also a jump in the density at $20\kpc$ which is associated
with the weak shock.  These features are discussed in more detail by
\citet{Graham08}.

Although the metallicity is not as well-constrained as the density or
temperature, the fluctuations in the radial profile were found to be
stable for different radial binning.  The structure in the metallicity
profile likely relates to real blobs of high or low metallicity in the
ICM (see \citet{SandersFabian07} for detailed metallicity maps).

A multi-component fit to the \textsc{dsdeproj} deprojected spectra was only
possible for the high quality spectra from the Perseus cluster.  Each
of the deprojected spectra were fitted with a variable number of
thermal components, using an F-test to ensure that the addition of
each one was statistically significant.  An F-test probability of less
than 0.1 was required for an extra model component to be added.  For
each component, the Galactic absorption and redshift were fixed to the
values in Table \ref{tab:obs} and the abundance parameters were tied
together.  The temperature, model normalization and abundance parameters were
allowed to vary. 

Half of the radial bins required an extra thermal component at either
a lower temperature of $0.5-1\keV$ or a much higher temperature
component of $>10\keV$.  The northern edge of the sector includes a
section of the filamentary structures seen in the soft X-ray
(\citealt{FabianPer03}).  These filaments have been found to correlate
with detections of cool gas in H$\alpha$ (\citealt{Conselice01}) and CO
(\citealt{Salome06}).  Each of the deprojected annuli likely contains
a component of this cool gas but in differing proportions.
  
We found that the higher temperature component could be equally as
well fitted by a power-law component with a photon index $\Gamma\sim1.5$.
\citet{Sanders04} found evidence for a distributed hard emission
component surrounding the core of the Perseus cluster by fitting a
high-temperature thermal component.  If the origin is hot thermal gas,
this material could be from a shock caused by a merger with the nearby
high velocity system (\citealt{SandersFabian07}).  

\subsection{Abell 262}

\begin{figure}
  \centering
  \includegraphics[width=\columnwidth]{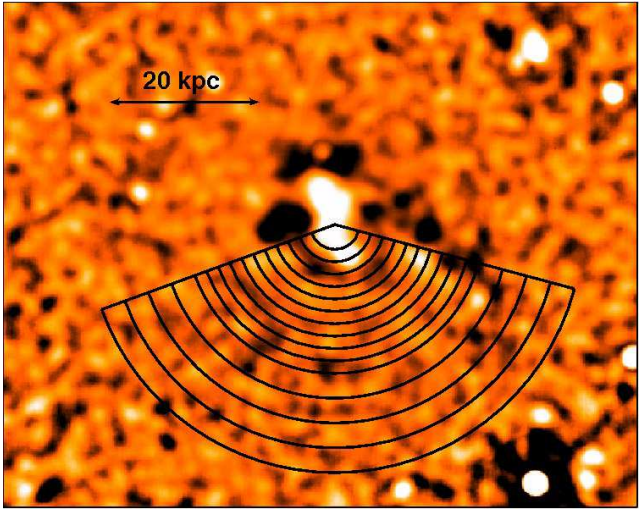}
  \caption{Unsharp mask image of Abell 262 made from the $0.3-7\keV$ band by
    subtracting an image smoothed with a Gaussian of width $10\asec$
    from one smoothed by $2.5\asec$ and dividing by the sum of the two
    images.  The annuli used for deprojection have been overlaid in black.}
  \label{fig:A262image}
\end{figure}

The inner regions of Abell 262 also host complex structures, including
bright knots of emission and a clear cavity to the east of the cluster
centre which is coincident with a lobe of radio emission
(\citealt{Blanton04}).  \citet{Blanton04} found that the knots of
structure in the core are located in the same region as optical
[N~\textsc{ii}] line emission, suggesting that the gas is cooling
down to temperatures of $\sim10^4\K$ in these regions.  

For a deprojection analysis, we selected a sector of the cluster that
covers the extended region of emission to the south of the nucleus and
avoids the $\sim5\kpc$ diameter bubble cavity (Figure \ref{fig:A262image}).  The effect of cavities
on the deprojection was examined in detail for Hydra A, which has a
much more complex system of bubbles.  For Abell 262, we focused on the
effect of the cool knots of emission.

\begin{figure}
  \includegraphics[width=\columnwidth]{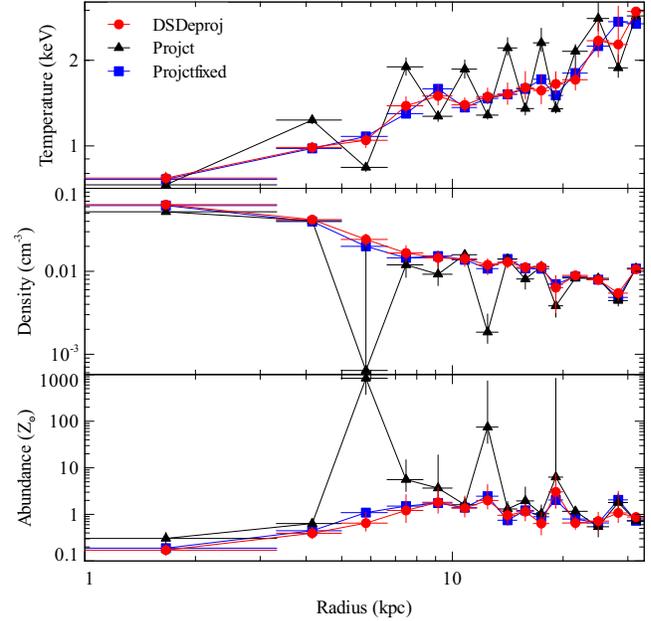}
  \caption{Deprojected temperature (top), electron density (centre)
    and metallicity (bottom) profiles for Abell 262.  There are no
    vertical error bars for the \textsc{projctfixed} model (blue line).}
  \label{fig:A262_DSDeproj}
\end{figure}

The single-temperature deprojection of this sector is shown in Figure
\ref{fig:A262_DSDeproj}.  \textsc{projct} generated an oscillating
temperature profile and a sharp discontinuity in the density and
abundance at $6\kpc$.  This is the only cluster in our sample for
which \textsc{projct} produced spurious results in all three parameter
profiles.  The discontinuity at $6\kpc$ coincides with the edge of the
dense core and the jump at $\sim13\kpc$ may correspond to a bright
knot of emission at the northern edge of the sector (Figure
\ref{fig:A262image}).

The sharp drop in the density parameter was caused by the unphysically
large value of the metallicity; the spectrum was dominated by line
emission with only a minimal amount of continuum.  We attempted to
solve this problem by placing constraints on the metallicity
parameter.  The \textsc{projct} deprojection was repeated with the
metallicity parameter in the third annulus, centred on $\sim6\kpc$, fixed to
$1\Zsun$.  However, this merely pushed the discontinuity into a
nearby annulus.  Imposing an upper limit of twice solar on the
metallicity parameter in each annulus (\citealt{Johnstone05}) reduced
the drop in density in the third annulus, although this was still
inconsistent with \textsc{dsdeproj} and the metallicity parameter
reached the allowed maximum of twice solar in seven annuli.  The
constraints on metallicity still produced oscillating temperature profiles.

\textsc{dsdeproj} produced a steady decline in temperature and a
smooth density profile with a steeper gradient around $6\kpc$
indicating the edge of the dense core.  We tested for the significance
of a second temperature component in the annulus centred on $6\kpc$ by
fitting the deprojected spectrum with a two-temperature model.
Although the density and abundance parameters were poorly constrained
in this model, the F-test probability indicated that the addition of a
second component at a temperature of $0.79^{+0.07}_{-0.06}\keV$
significantly improved the fit.

\subsection{Hydra A}

\begin{figure}
  \centering
  \includegraphics[width=\columnwidth]{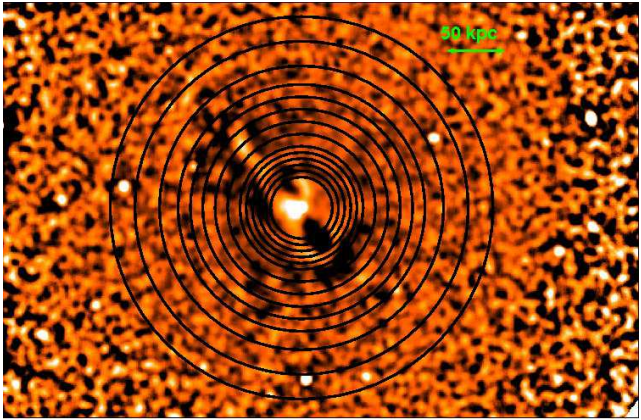}
  \caption{Unsharp mask image of Hydra A made from the $0.3-7\keV$
    band by subtracting an image smoothed with a Gaussian of width
    $10\asec$ from one smoothed by $2.5\asec$ and dividing by the sum
    of the two images.  The annuli used for deprojection have been
    overlaid in black, excluding the inner 9 annuli which are very
    finely spaced.  The innermost black circle corresponds to a radius
    of $26\kpc$.}
  \label{fig:Hydraimage}
\end{figure}

Hydra A (Abell 780) was observed by \emph{Chandra} for two pointings
of $100\ks$ in ACIS-S (ObsID 4969 and 4970).  Spectra were extracted
from complete annuli in both observations, deprojected and fitted
together in \textsc{xspec}.  The central $1.5\asec$ of the cluster,
containing the AGN, was excluded from the regions analysed (Figure \ref{fig:Hydraimage}). 

The large-scale X-ray morphology of this cluster is generally smooth,
however the core reveals the presence of cavities in the ambient
cluster gas caused by outbursts from the central AGN
(\citealt{McNamara00}; \citealt{David01}; \citealt{Nulsen02};
\citealt{Nulsen05}; \citealt{Wise07}).  These bubbles in the ICM are
coincident with radio lobes from the central FR type I radio source 3C
218 (\citealt{McNamara00}).  \citet{Nulsen02} found that cooler gas
extends outward from the centre of the cluster, beyond the cavities,
in the direction of the radio source axis.  There will therefore be
significant quantities of cooler gas in each deprojected annulus.

\begin{figure}
  \includegraphics[width=\columnwidth]{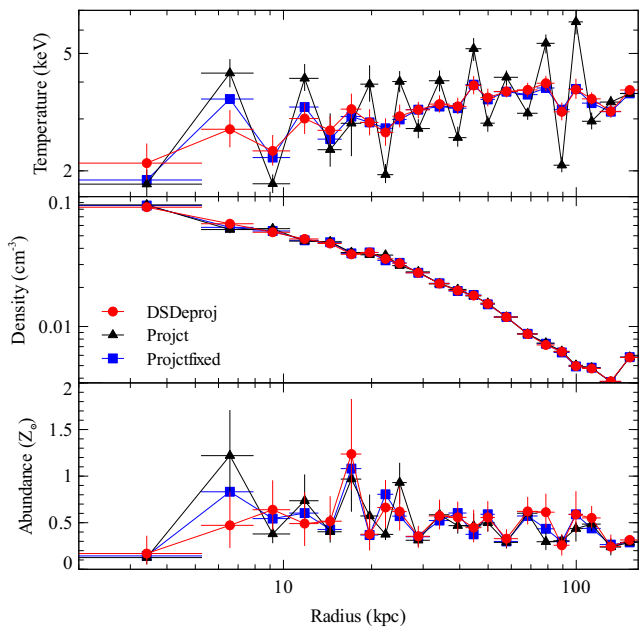}
  \caption{Deprojected temperature (top), electron density (centre)
    and metallicity (bottom) profiles for Hydra A.  There are no
    vertical error bars for the \textsc{projctfixed} model (blue line).}
  \label{fig:Hydra_DSDeproj}
\end{figure}

The single-temperature deprojection of Hydra A is shown in Figure
\ref{fig:Hydra_DSDeproj}.  The \textsc{projct} temperature profile for
this cluster is particularly unstable with large oscillations in the
temperature parameter throughout the core.  Although the \textsc{dsdeproj}
result also appears to oscillate, the errors are consistent with a
smooth temperature profile for the inner radial
bins.  The density profiles show a smooth increase towards the centre
of the cluster with no features that can be readily associated with
the large system of cavities in the X-ray emission.  The effects of
this structure were averaged out when using complete annuli.  The
metallicity profile was approximately consistent with a constant
value of $\sim0.5\Zsun$.

Two sectors of Hydra A, one containing the northern set of cavities
and the second about a perpendicular axis, were separately
deprojected but we were unable to significantly detect the inner X-ray
cavities in the former.


\section{\textsc{dsdeproj} Validation}
\label{validation}
Multiphase gas in observed clusters is not evenly distributed but
tends to be concentrated in cluster cores and, in addition, the ICM
contains sharp features such as shocks and cold fronts.  We have
tested \textsc{dsdeproj} on a variety of simulated clusters containing
examples of these features to check that the expected result was
recovered.  

\begin{figure}
  \includegraphics[width=\columnwidth]{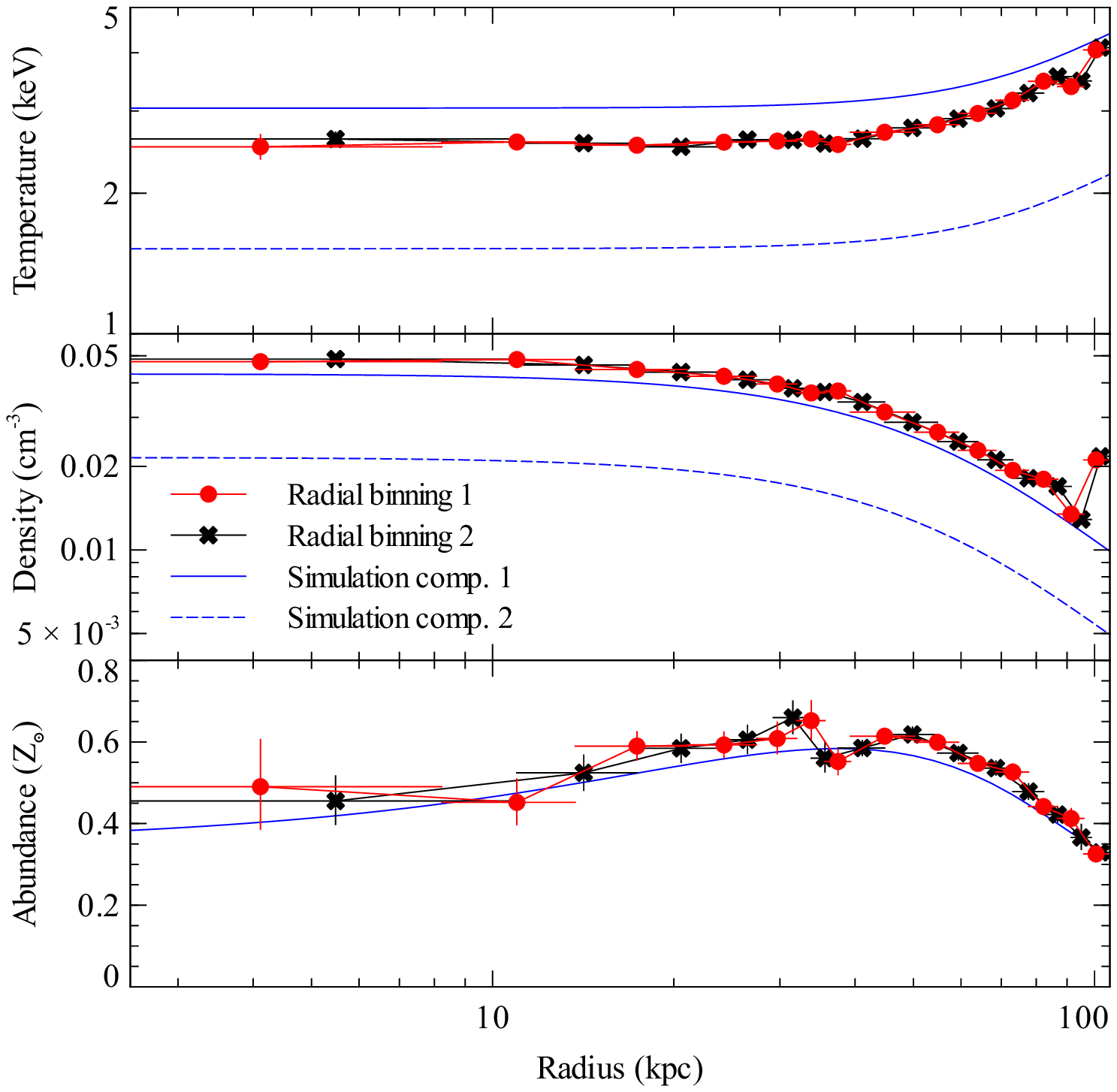}
  \caption{Deprojected temperature (top), electron density (centre)
    and metallicity (bottom) profiles for a two-temperature simulated
    cluster.  The single-temperature \textsc{dsdeproj} results for two
    different radial bin distributions are overlaid on the true
    cluster profiles.}
  \label{fig:Fakecluster_binning}
\end{figure}

\begin{figure}
  \includegraphics[width=\columnwidth]{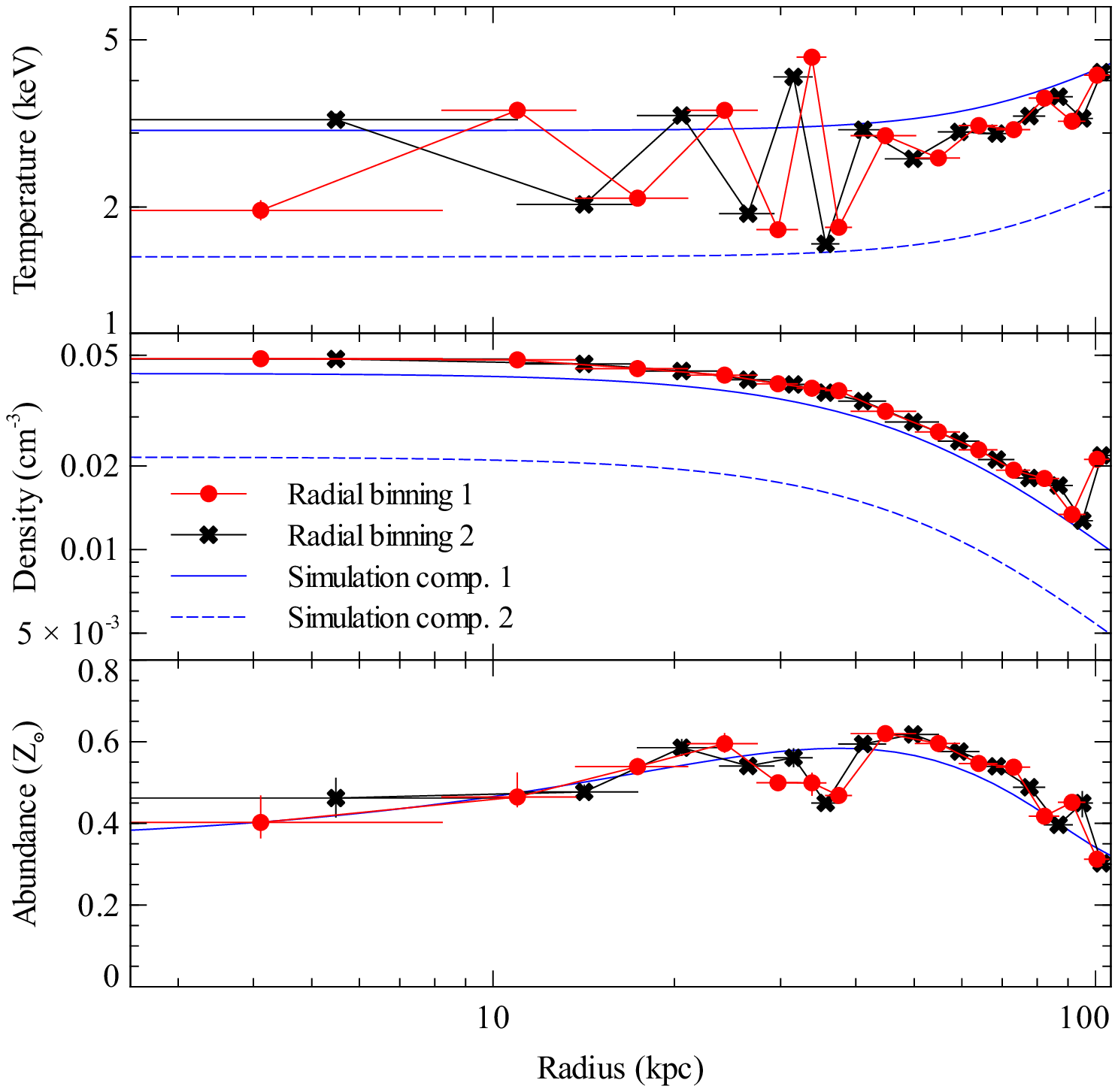}
  \caption{Deprojected temperature (top), electron density (centre)
    and metallicity (bottom) profiles for a two-temperature simulated
    cluster.  The single-temperature \textsc{projct} results for two
    different radial bin distributions are overlaid on the true
    cluster profiles.}
  \label{fig:Fakecluster_binning_projct}
\end{figure}

\textsc{dsdeproj} calculates deprojected profiles that are stable for
almost any choice of radial binning.  Figure
\ref{fig:Fakecluster_binning} shows the \textsc{dsdeproj} deprojection
of a simulated galaxy cluster under two different radial bin
distributions.  The two \textsc{dsdeproj} deprojections produce
profiles that are consistent and stable.  As we show later,
instabilities only occur when a radial bin contains drastically
different gas properties, such as both sides of shock.  For comparison, we show the
equivalent result under a deprojection with \textsc{projct} in Figure
\ref{fig:Fakecluster_binning_projct}.

\begin{figure}
  \includegraphics[width=\columnwidth]{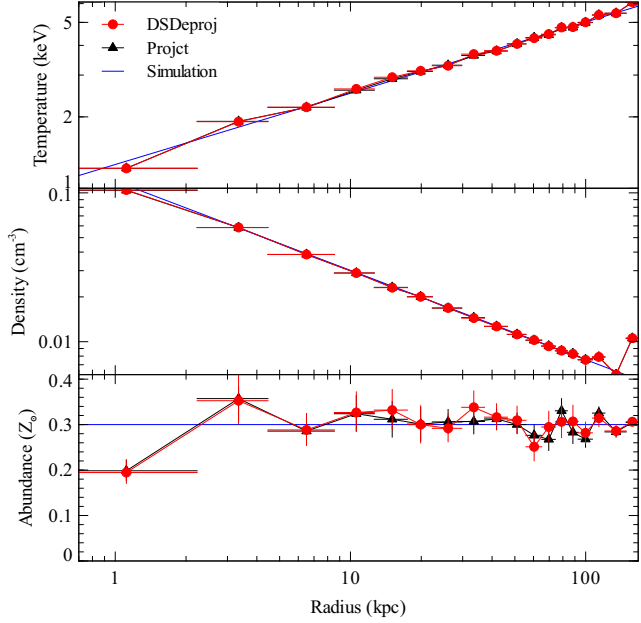}
  \caption{Deprojected temperature (top), electron density (centre)
    and metallicity (bottom) profiles for a single-temperature
    simulated cluster with a steep power-law density profile.}
  \label{fig:Fakecluster_steep}
\end{figure}

In section \ref{sec:DSDeproj}, we use a simulated cluster based on the
Perseus cluster to show that \textsc{dsdeproj} produces smooth, stable
temperature profiles which accurately reproduce the simulated
profile.  The density and temperature profiles for these simulated
clusters are particularly flat in the radial region considered so
we also tested \textsc{dsdeproj} on simulated clusters
with steep power-law profiles in temperature and density
given by the equations 

\begin{align}
\nonumber T &= 1.26r^{0.3} \keV \\
 &= 12 \keV \;\;\; \mathrm{for} \: r > 2\Mpc
\end{align}
\begin{equation}
n_e=0.12\left(\frac{r^{-0.6}}{\kpc}\right) \pcmcu
\end{equation}
\begin{equation}
Z = 0.3\Zsun
\end{equation}

\noindent where $r$ is the cluster radius in units of kpc.  Both
\textsc{dsdeproj} and \textsc{projct} correctly reproduce the steep
temperature drop and density profile (Figure
\ref{fig:Fakecluster_steep}).  

\begin{figure}
  \includegraphics[width=\columnwidth]{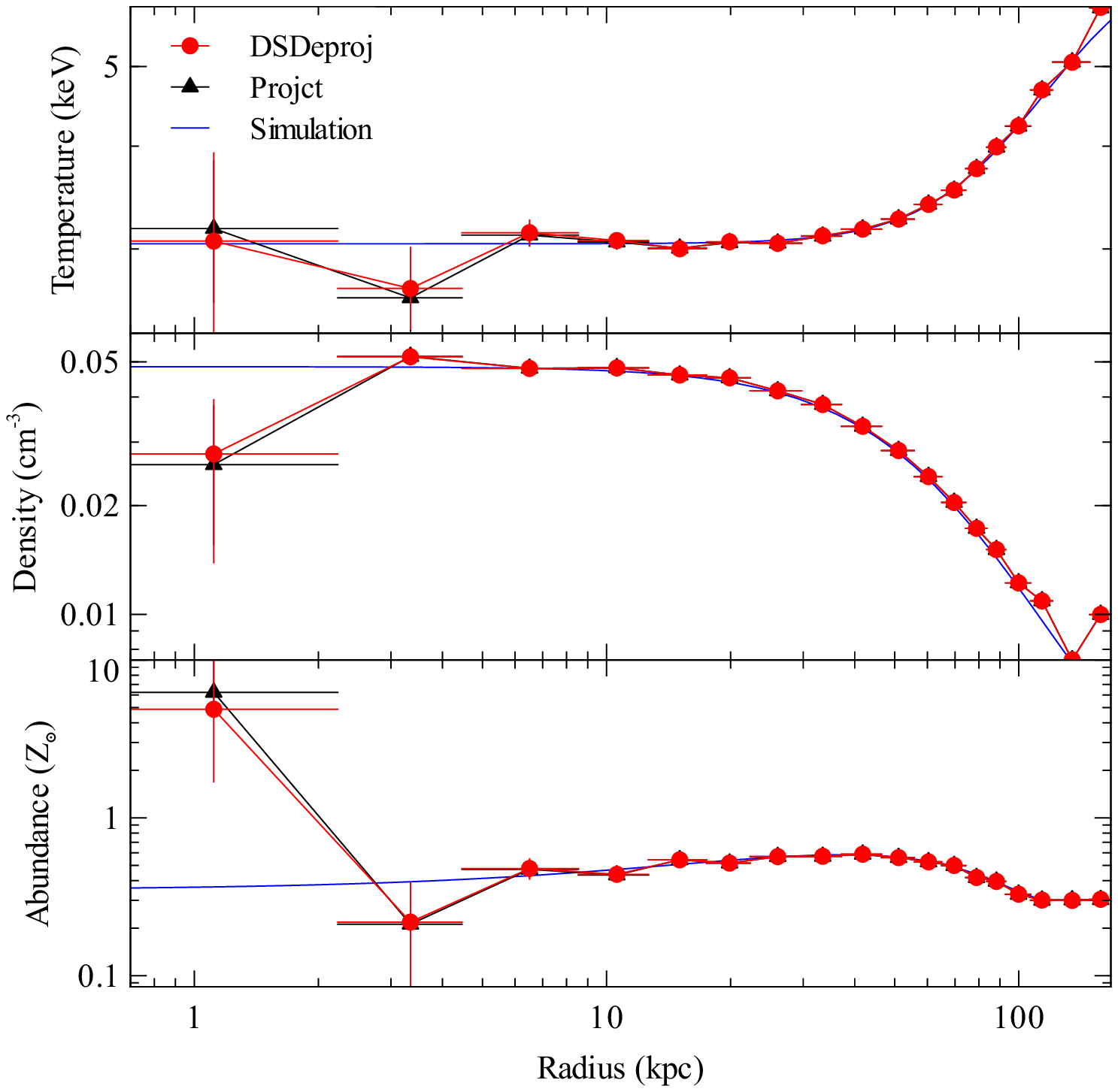}
  \caption{Deprojected temperature (top), electron density (centre)
    and metallicity (bottom) profiles for a single-temperature
    simulated cluster which is elongated along the line of sight.}
  \label{fig:Fakecluster_stretched}
\end{figure}

We tested the assumption of spherical
symmetry by applying \textsc{dsdeproj} to a cluster that was stretched
by a third along the line of sight (Figure
\ref{fig:Fakecluster_stretched}).  \textsc{dsdeproj} and
\textsc{projct} correctly reproduce the expected profiles; the central
radial bin is poorly constrained because the spectrum contains
residuals from the incorrect subtraction of the outer layers.
Assuming incorrect cluster geometry does not produce oscillating
profiles in \textsc{projct} (or \textsc{dsdeproj}), however
\textsc{dsdeproj} could be modified for the deprojection of
non-spherical systems.

\begin{figure}
  \includegraphics[width=\columnwidth]{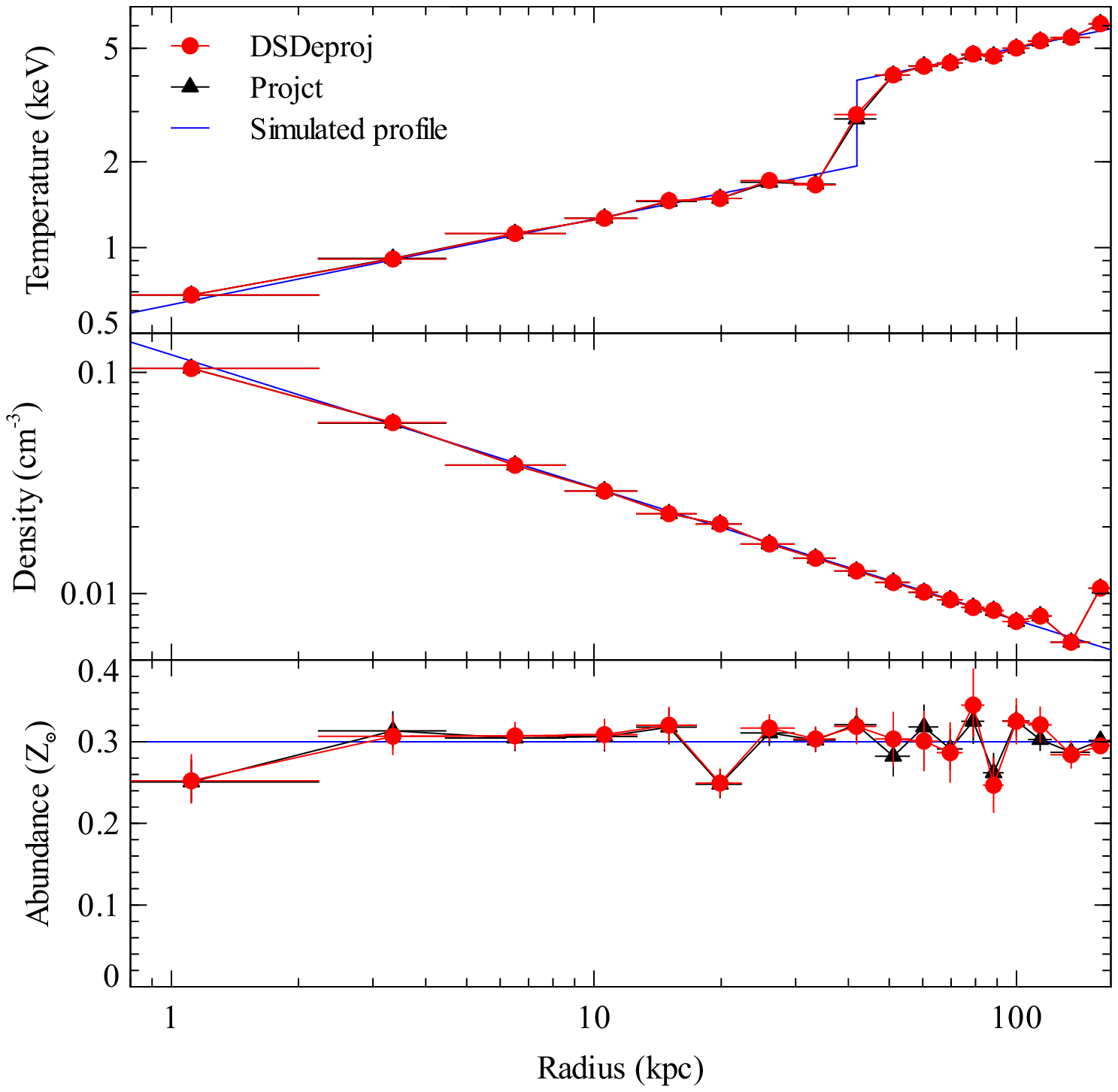}
  \caption{Deprojected temperature (top), electron density (centre)
    and metallicity (bottom) profiles for a single-temperature
    simulated cluster which has a sharp temperature break.}
  \label{fig:Fakecluster_break}
\end{figure}

\begin{figure}
  \includegraphics[width=\columnwidth]{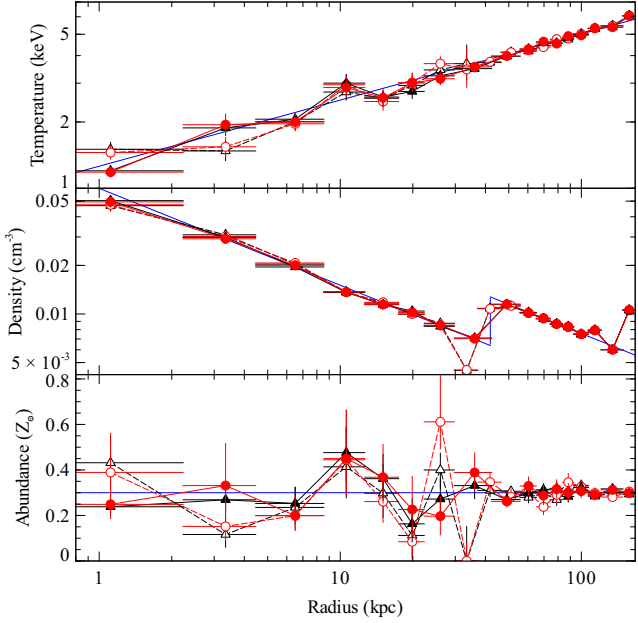}
  \caption{Deprojected temperature (top), electron density (centre)
    and metallicity (bottom) profiles for a single-temperature
    simulated cluster which has a sharp density break.  The true
    cluster profile is shown by the blue line; the \textsc{dsdeproj}
    and \textsc{projct} deprojected profiles are shown by red circles
    and black triangles respectively.  The open points with dashed
    lines indicate a deprojection where the density jump was
    positioned in the centre of a radial bin; solid points and lines
    denote the deprojection with the density jump shifted to the edge
    of the radial bin.}
  \label{fig:Fakecluster_breakdens}
\end{figure}

We also constructed clusters with a sharp drop or rise in temperature
or density at a particular radius.  Figure \ref{fig:Fakecluster_break}
shows the deprojected profiles of a cluster with a temperature break
at $42\kpc$.  \textsc{dsdeproj} and \textsc{projct} both reproduce the breaks
in temperature without any additional oscillation in the deprojected
profile.  However sharp changes in density can cause additional
ringing in the density and temperature profiles (dashed lines in
Figure \ref{fig:Fakecluster_breakdens}).  The density parameter in the
shell containing the density break cannot account for the two distinct
values causing an under- or overestimation of the projection onto the
next shell.  This problem can be alleviated by shifting the radial
bins so that the density jump occurs close to the edge of a bin (solid
lines in Figure \ref{fig:Fakecluster_breakdens}).  

\begin{figure}
  \includegraphics[width=\columnwidth]{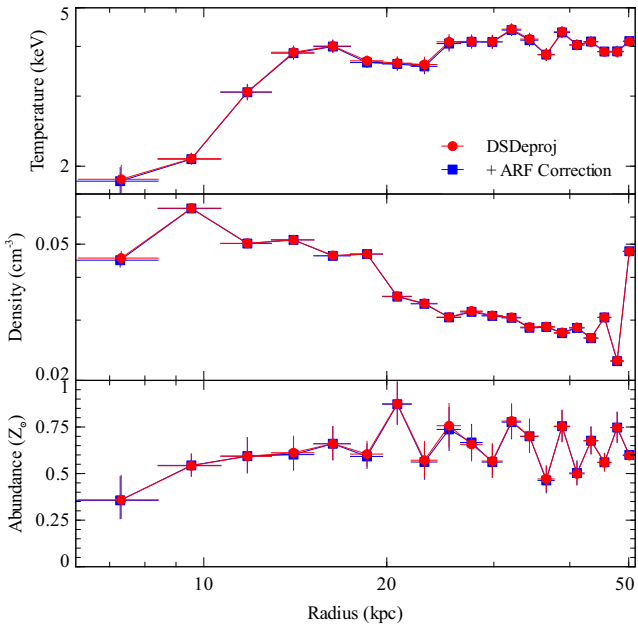}
  \caption{Deprojected temperature (top), electron density (centre)
    and metallicity (bottom) profiles for the Perseus cluster.  The
    deprojected profiles were created with (blue line) and without
    (red line) an ancillary response correction to illustrate that
    there is no significant difference.}
  \label{fig:Perseus_arf}
\end{figure}

As a final note, \textsc{dsdeproj} assumes that the response of the detector
does not vary significantly, which is the case for the \emph{Chandra}
ACIS-S3 used in this work.  An additional routine which uses the
ancillary response files to correct for changes in the effective area
has been incorporated into \textsc{dsdeproj}, however this is not required
here.  For the ACIS-S3 detector any variation in effective area has a
negligible effect on the deprojection (Figure \ref{fig:Perseus_arf}).

\section{Summary}
We have investigated several issues regarding the reliable
deprojection of galaxy clusters and validated the deprojection
routine \textsc{dsdeproj}.  Deprojection methods that assume a
spectral model for the deprojection, such as \textsc{projct} in
\textsc{xspec}, have previously been found to produce temperature
profiles which bounce between unphysical values separated by several
times their uncertainty.  We have shown that this effect is caused by
fitting multiphase gas with a single-temperature model.  Although a
two-temperature \textsc{projct} deprojection recovers the true
profiles for the simulated clusters, the majority of the cluster
observations currently available in the \emph{Chandra} archive do not
have a sufficient number of counts to constrain the two-components in
a detailed deprojection.

\textsc{dsdeproj}, our deprojection routine which assumes only spherical
geometry, solves some of these issues inherent to model-dependent
deprojection routines.  \textsc{dsdeproj} produces a set of `deprojected
spectra' which can then be fitted by a suitable spectral model in
\textsc{xspec}.  We have shown that this method does not generate the
oscillating temperature profiles for multi-temperature clusters and
produces a stable solution for an elongated cluster and clusters with
breaks in temperature or density. 

\textsc{projct} and \textsc{dsdeproj} were applied to a small sample of three
nearby galaxy clusters, each of which has a component of cooler gas or
complex substructure in the core.  \textsc{projct} produced a rapidly
oscillating temperature profile for each cluster and, in Abell 262,
unphysical values of density and abundance.  The deprojected profiles
generated by \textsc{dsdeproj} were smoothly varying, producing stable
solutions at shocks, knots of substructure and with different radial
binning.  \textsc{dsdeproj} was only able to constrain a two-temperature
spectral model for the Perseus cluster deprojection.  This revealed a
low temperature component in several annuli that contain cool gas
filaments and a power-law component associated with a distributed hard
emission component (\citealt{Sanders04}).  

The \textsc{dsdeproj} source code is available at www-xray.ast.cam.ac.uk/papers/dsdeproj

\section*{Acknowledgments}
HRR and ACF acknowledge support from the Science and Technology
Facilities Council and the Royal Society, respectively.  We thank the
referee for helpful comments.

\bibliographystyle{mnras}
\bibliography{refs.bib}

\begin{thebibliography}{}

\bibitem[\protect\citeauthoryear{{Anders} \& {Grevesse}}{{Anders} \&
  {Grevesse}}{1989}]{AndersGrevesse89}
{Anders} E.,  {Grevesse} N., 1989, \gca, 53, 197

\bibitem[\protect\citeauthoryear{{Arnaud}}{{Arnaud}}{1996}]{Arnaud96}
{Arnaud} K.~A., 1996, in Astronomical Society of the Pacific Conference Series,
  Vol. 101, {Jacoby} G.~H.,  {Barnes} J., ed, Astronomical Data Analysis
  Software and Systems V, p.~17

\bibitem[\protect\citeauthoryear{{Balucinska-Church} \&
  {McCammon}}{{Balucinska-Church} \& {McCammon}}{1992}]{Balucinska92}
{Balucinska-Church} M.,  {McCammon} D., 1992, \apj, 400, 699

\bibitem[\protect\citeauthoryear{{B{\^i}rzan} et~al.}{{B{\^i}rzan}
  et~al.}{2004}]{Birzan04}
{B{\^i}rzan} L., {Rafferty} D.~A., {McNamara} B.~R., {Wise} M.~W.,  {Nulsen}
  P.~E.~J., 2004, \apj, 607, 800

\bibitem[\protect\citeauthoryear{{Blanton} et~al.}{{Blanton}
  et~al.}{2004}]{Blanton04}
{Blanton} E.~L., {Sarazin} C.~L., {McNamara} B.~R.,  {Clarke} T.~E., 2004,
  \apj, 612, 817

\bibitem[\protect\citeauthoryear{{Boehringer} et~al.}{{Boehringer}
  et~al.}{1993}]{Boehringer93}
{Boehringer} H., {Voges} W., {Fabian} A.~C., {Edge} A.~C.,  {Neumann} D.~M.,
  1993, \mnras, 264, L25

\bibitem[\protect\citeauthoryear{{Churazov} et~al.}{{Churazov}
  et~al.}{2003}]{Churazov03}
{Churazov} E., {Forman} W., {Jones} C.,  {B{\"o}hringer} H., 2003, \apj, 590,
  225

\bibitem[\protect\citeauthoryear{{Conselice}, {Gallagher}, \&
  {Wyse}}{{Conselice} et~al.}{2001}]{Conselice01}
{Conselice} C.~J., {Gallagher} J.~S., III,  {Wyse} R.~F.~G., 2001, \aj, 122,
  2281

\bibitem[\protect\citeauthoryear{{David} et~al.}{{David}
  et~al.}{2001}]{David01}
{David} L.~P., {Nulsen} P.~E.~J., {McNamara} B.~R., {Forman} W., {Jones} C.,
  {Ponman} T., {Robertson} B.,  {Wise} M., 2001, \apj, 557, 546

\bibitem[\protect\citeauthoryear{{Dunn} \& {Fabian}}{{Dunn} \&
  {Fabian}}{2006}]{DunnFabian06}
{Dunn} R.~J.~H.,  {Fabian} A.~C., 2006, \mnras, 373, 959

\bibitem[\protect\citeauthoryear{{Edge} \& {Frayer}}{{Edge} \&
  {Frayer}}{2003}]{EdgeFrayer03}
{Edge} A.~C.,  {Frayer} D.~T., 2003, \apjl, 594, L13

\bibitem[\protect\citeauthoryear{{Ettori} \& {Fabian}}{{Ettori} \&
  {Fabian}}{2000}]{Ettori00}
{Ettori} S.,  {Fabian} A.~C., 2000, \mnras, 317, L57

\bibitem[\protect\citeauthoryear{{Fabian}}{{Fabian}}{1994}]{Fabian94}
{Fabian} A.~C., 1994, \araa, 32, 277

\bibitem[\protect\citeauthoryear{{Fabian} et~al.}{{Fabian}
  et~al.}{2003}]{FabianPer03}
{Fabian} A.~C., {Sanders} J.~S., {Allen} S.~W., {Crawford} C.~S., {Iwasawa} K.,
  {Johnstone} R.~M., {Schmidt} R.~W.,  {Taylor} G.~B., 2003, \mnras, 344, L43

\bibitem[\protect\citeauthoryear{{Fabian} et~al.}{{Fabian}
  et~al.}{2000}]{FabianPer00}
{Fabian} A.~C. et~al., 2000, \mnras, 318, L65

\bibitem[\protect\citeauthoryear{{Fabian} et~al.}{{Fabian}
  et~al.}{2006}]{FabianPer06}
{Fabian} A.~C., {Sanders} J.~S., {Taylor} G.~B., {Allen} S.~W., {Crawford}
  C.~S., {Johnstone} R.~M.,  {Iwasawa} K., 2006, \mnras, 366, 417

\bibitem[\protect\citeauthoryear{{Forman} et~al.}{{Forman}
  et~al.}{2005}]{FormanM8705}
{Forman} W. et~al., 2005, \apj, 635, 894

\bibitem[\protect\citeauthoryear{{Graham}, {Fabian}, \& {Sanders}}{{Graham}
  et~al.}{2008}]{Graham08}
{Graham} J., {Fabian} A.~C.,  {Sanders} J.~S., 2008, \mnras, 349

\bibitem[\protect\citeauthoryear{{Hicks} \& {Mushotzky}}{{Hicks} \&
  {Mushotzky}}{2005}]{HicksMushotzky05}
{Hicks} A.~K.,  {Mushotzky} R., 2005, \apjl, 635, L9

\bibitem[\protect\citeauthoryear{{Johnstone} et~al.}{{Johnstone}
  et~al.}{2005}]{Johnstone05}
{Johnstone} R.~M., {Fabian} A.~C., {Morris} R.~G.,  {Taylor} G.~B., 2005,
  \mnras, 356, 237

\bibitem[\protect\citeauthoryear{{Johnstone}, {Fabian}, \&
  {Nulsen}}{{Johnstone} et~al.}{1987}]{Johnstone87}
{Johnstone} R.~M., {Fabian} A.~C.,  {Nulsen} P.~E.~J., 1987, \mnras, 224, 75

\bibitem[\protect\citeauthoryear{{Jones} \& {Forman}}{{Jones} \&
  {Forman}}{1999}]{JonesForman99}
{Jones} C.,  {Forman} W., 1999, \apj, 511, 65

\bibitem[\protect\citeauthoryear{{Kaastra}}{{Kaastra}}{1992}]{Kaastra92}
{Kaastra} J.~S., 1992, in Internal SRON-Leiden Report, updated version 2.0

\bibitem[\protect\citeauthoryear{{Kaastra} et~al.}{{Kaastra}
  et~al.}{2004}]{Kaastra04}
{Kaastra} J.~S. et~al., 2004, \aap, 413, 415

\bibitem[\protect\citeauthoryear{{Kalberla} et~al.}{{Kalberla}
  et~al.}{2005}]{Kalberla05}
{Kalberla} P.~M.~W., {Burton} W.~B., {Hartmann} D., {Arnal} E.~M., {Bajaja} E.,
  {Morras} R.,  {P{\"o}ppel} W.~G.~L., 2005, \aap, 440, 775

\bibitem[\protect\citeauthoryear{{Kriss}, {Cioffi}, \& {Canizares}}{{Kriss}
  et~al.}{1983}]{Kriss83}
{Kriss} G.~A., {Cioffi} D.~F.,  {Canizares} C.~R., 1983, \apj, 272, 439

\bibitem[\protect\citeauthoryear{{Liedahl}, {Osterheld}, \&
  {Goldstein}}{{Liedahl} et~al.}{1995}]{Liedahl95}
{Liedahl} D.~A., {Osterheld} A.~L.,  {Goldstein} W.~H., 1995, \apjl, 438, L115

\bibitem[\protect\citeauthoryear{{McNamara} \& {Nulsen}}{{McNamara} \&
  {Nulsen}}{2007}]{McNamaraNulsen07}
{McNamara} B.~R.,  {Nulsen} P.~E.~J., 2007, \araa, 45, 117

\bibitem[\protect\citeauthoryear{{McNamara} et~al.}{{McNamara}
  et~al.}{2006}]{McNamara06}
{McNamara} B.~R. et~al., 2006, \apj, 648, 164

\bibitem[\protect\citeauthoryear{{McNamara} et~al.}{{McNamara}
  et~al.}{2000}]{McNamara00}
{McNamara} B.~R. et~al., 2000, \apjl, 534, L135

\bibitem[\protect\citeauthoryear{{Mewe}, {Gronenschild}, \& {van den
  Oord}}{{Mewe} et~al.}{1985}]{Mewe85}
{Mewe} R., {Gronenschild} E.~H.~B.~M.,  {van den Oord} G.~H.~J., 1985, \aaps,
  62, 197

\bibitem[\protect\citeauthoryear{{Mewe}, {Lemen}, \& {van den Oord}}{{Mewe}
  et~al.}{1986}]{Mewe86}
{Mewe} R., {Lemen} J.~R.,  {van den Oord} G.~H.~J., 1986, \aaps, 65, 511

\bibitem[\protect\citeauthoryear{{Nulsen} \& {Bohringer}}{{Nulsen} \&
  {Bohringer}}{1995}]{Nulsen95}
{Nulsen} P.~E.~J.,  {Bohringer} H., 1995, \mnras, 274, 1093

\bibitem[\protect\citeauthoryear{{Nulsen} et~al.}{{Nulsen}
  et~al.}{2002}]{Nulsen02}
{Nulsen} P.~E.~J., {David} L.~P., {McNamara} B.~R., {Jones} C., {Forman} W.~R.,
   {Wise} M., 2002, \apj, 568, 163

\bibitem[\protect\citeauthoryear{{Nulsen} et~al.}{{Nulsen}
  et~al.}{2005}]{Nulsen05}
{Nulsen} P.~E.~J., {McNamara} B.~R., {Wise} M.~W.,  {David} L.~P., 2005, \apj,
  628, 629

\bibitem[\protect\citeauthoryear{{O'Dea} et~al.}{{O'Dea} et~al.}{2008}]{ODea08}
{O'Dea} C.~P. et~al., 2008, astro-ph/0803.1772

\bibitem[\protect\citeauthoryear{{Peterson} \& {Fabian}}{{Peterson} \&
  {Fabian}}{2006}]{PetersonFabian06}
{Peterson} J.~R.,  {Fabian} A.~C., 2006, \physrep, 427, 1

\bibitem[\protect\citeauthoryear{{Peterson} et~al.}{{Peterson}
  et~al.}{2003}]{Peterson03}
{Peterson} J.~R., {Kahn} S.~M., {Paerels} F.~B.~S., {Kaastra} J.~S., {Tamura}
  T., {Bleeker} J.~A.~M., {Ferrigno} C.,  {Jernigan} J.~G., 2003, \apj, 590,
  207

\bibitem[\protect\citeauthoryear{{Rafferty} et~al.}{{Rafferty}
  et~al.}{2006}]{Rafferty06}
{Rafferty} D.~A., {McNamara} B.~R., {Nulsen} P.~E.~J.,  {Wise} M.~W., 2006,
  \apj, 652, 216

\bibitem[\protect\citeauthoryear{{Salom{\'e}} et~al.}{{Salom{\'e}}
  et~al.}{2006}]{Salome06}
{Salom{\'e}} P. et~al., 2006, \aap, 454, 437

\bibitem[\protect\citeauthoryear{{Sanders} \& {Fabian}}{{Sanders} \&
  {Fabian}}{2007}]{SandersFabian07}
{Sanders} J.~S.,  {Fabian} A.~C., 2007, \mnras, 381, 1381

\bibitem[\protect\citeauthoryear{{Sanders} et~al.}{{Sanders}
  et~al.}{2004}]{Sanders04}
{Sanders} J.~S., {Fabian} A.~C., {Allen} S.~W.,  {Schmidt} R.~W., 2004, \mnras,
  349, 952

\bibitem[\protect\citeauthoryear{{Vikhlinin}, {Markevitch}, \&
  {Murray}}{{Vikhlinin} et~al.}{2001}]{Vikhlinin01}
{Vikhlinin} A., {Markevitch} M.,  {Murray} S.~S., 2001, \apj, 551, 160

\bibitem[\protect\citeauthoryear{{Wise} et~al.}{{Wise} et~al.}{2007}]{Wise07}
{Wise} M.~W., {McNamara} B.~R., {Nulsen} P.~E.~J., {Houck} J.~C.,  {David}
  L.~P., 2007, \apj, 659, 1153

\end{thebibliography}

\clearpage

\end{document}